\documentclass{ckm}            

\newcommand{\nn}{\nonumber}
\newcommand{\bea}{\begin{eqnarray}}
\newcommand{\eea}{\end{eqnarray}}

\def \beq{\begin{equation}}
\def \eeq{\end{equation}}

\def \Im{{\hbox{Im}}\,}

\def \cl#1{{#1\%\ \mathrm{C.L.}}}

\def \fig#1{Fig.~\ref{#1}}

\def \nn{\nonumber}
\def \rf{Ref.~\cite}

\def \bit{\begin{itemize}}
\def \eit{\end{itemize}}

\def \b{\beta}
\def \D{\Delta}
\def \g{\gamma}

\def \d{\delta}

\def \l{\lambda}

\def \m{\mu}

\def \r{\rho}

\def \z{\zeta}
\def \et{\eta}

\confname{Workshop on the CKM Unitarity Triangle, IPPP Durham, April
  2003}

\title{On Exclusive and Inclusive Rare $B$ Decays: \\ 
   CKM Phenomenology and New Physics Reach \thanks{based on invited 
talks given 
by T.H. and E.L. at the Workshop on the CKM Unitarity Triangle, IPPP Durham,
England, April 2003, and on an invited 
talk given by T.H. at the Ringberg Phenomenology 
Workshop on Heavy Flavours, Ringberg Castle, 
Germany, May 2003. \mbox{CERN-TH/2003-144, \, SLAC-PUB-10042,\, ZU-TH-12/03}}}

\author{Tobias Hurth\thanks{Heisenberg Fellow}\addressmark{a}
 and Enrico Lunghi\addressmark{c} }
\address[a]{CERN, Theory Division, CH-1211 Geneva 23, Switzerland,\\
SLAC, Stanford University, Stanford, CA 94309, USA}
\address[b]{Institute for Theoretical Physics, University of Zurich,\\
CH-8057 Zurich, Switzerland}

\begin{document}

\begin{abstract}
  We report recent results in rare $B$ decays.  Our focus will be on
  $b\to d \gamma$ and $b \to s \ell^+ \ell^-$ transitions. We discuss
  their impact on the CKM phenomenology and their role within our
  search for new physics.  In particular, we analyse the impact of a
  recent lattice QCD estimate of the $B \rightarrow K^*$ form factor
  at zero recoil.  We also briefly discuss the presently available
  optimal theoretical tools for the inclusive and exclusive modes.

\end{abstract}

\maketitle


\section{Introduction}
Rare $B$ decays, as flavour changing neutral current processes (FCNC),
are the most important tools within our (indirect) search for new
physics in the present era of $B$ factories (for a recent review see
\cite{Hurth}).  Among them, the inclusive $b \rightarrow s \gamma$
mode is still the most prominent, because it has already measured by
several independent experiments
\cite{CLEOincl,CLEOneu,ALEPH,Bellebsg,Babarbsg}. The present world
average is \cite{Colin}: \beq
\label{world}
{\cal B}(B \to X_s \gamma) = (3.34 \pm 0.38) \times 10^{-4}.  \eeq The
present next-to-leading -logarithmic (NLL) QCD prediction, based on
the original QCD calculations of several groups
(\cite{AG91,GHW,Adel,Mikolaj}), has an additional charm mass
renormalization scheme ambiguity as was first noticed in
\cite{GambinoMisiak}. The $\overline{MS}$ scheme is used in the most
recent updates \cite{GambinoMisiak,Czarnecki}.  This choice is guided
by the experience gained from many higher order calculations where the
mass is dominantly off-shell and leads to the following theoretical
prediction for the `total' branching ratio:
\begin{equation} 
\label{totalbr}
{\cal B}(B \to X_s \gamma) = (3.70 \pm 0.30) \times 10^{-4}.
\end{equation}
However, the renormalization scheme for $m_c$ is an NNLL issue, and
one should regard the preference for the $\overline{MS}$ scheme in
comparison with the pole mass scheme (which was used within all
previous NLL predictions) just as an educated guess of the presently
unknown NNLL corrections. Therefore, one could also argue for a
slightly larger theoretical error in (\ref{totalbr}). A complete
resolution of this problem can be achieved by a NNLL calculation which
is presently under study.

The stringent bounds obtained from the $B \rightarrow X_s \gamma$ mode
on various non-standard scenarios (see
e.g.~\cite{Carena,Degrassinew,Giannew,OUR,NEWNEW}) are a clear example
of the importance of clean FCNC observables in discriminating
new-physics models.  The exclusive modes, however, often have large
uncertainties due to the hadronic form factors and are not clean
enough to disentangle possible new physics effects from hadronic
uncertainties, but they can serve as important QCD tests.  Exceptions
are ratios of exclusive quantities like asymmetries in which a large
part of the hadronic uncertainties cancel out.

Regarding the corresponding exclusive mode $B \to K^* \gamma$, quite
recently a preliminary lattice determination of the $B\to K^*$ form
factor at zero recoil~\cite{damir} was presented which is in perfect
agreement with the previous indirect determinations. We analyse the
impact of this new result on the CKM phenomenology in section 3.2.

Besides the $b \rightarrow s \gamma$ mode, also the $b \rightarrow s
\ell^+\ell^-$ transitions are already accessible at $B$ factories
\cite{BELLEbsll1,BABARbsll1,BELLEbsll2}, inclusively and exclusively.
The $b \rightarrow s \ell^+ \ell^-$ mode represents new sources of
interesting observables, particularly kinematic observables such as
the invariant dilepton mass spectrum and the forward--backward (FB)
asymmetry.  Rare $B$ decays are also relevant to the CKM
phenomenology; the $b \rightarrow d \gamma$ is especially important in
this respect.  In the following, we will focus on the latter two rare
modes and also briefly discuss the theoretical tools available for the
analysis of exclusive and inclusive channels.

\section{Theoretical Tools}
The effective field theory approach serves as a theoretical framework
for both inclusive and exclusive modes.  The standard method of the
operator product expansion (OPE) allows for a separation of the $B$
meson decay amplitude into two distinct parts, the long-distance
contributions contained in the operator matrix elements and the
short-distance physics described by the so-called Wilson coefficients.
The $W$ boson and the top quark with mass larger than the
factorization scale are integrated out, i.e. removed from the theory
as dynamical fields. The effective Hamiltonian for radiative and
semileptonic $b \rightarrow s/d$ transitions in the SM
can be written as 
\beq
    \label{Heff}
    {\cal H}_{eff} =  - \frac{4G_F}{\sqrt{2}}
    \left[ \l_q^t \sum_{i=1}^{10} C_i {\cal O}_i  + 
           \l_q^u\sum_{i=1}^{2} C_i  ({\cal O}_i  - {\cal O}_i^u) \right]
\eeq
where ${\cal O}_i(\mu)$ are dimension-six operators at the scale $\m
\sim O(m_b)$; $C_i(\mu)$ are the corresponding Wilson coefficients.
Clearly, only in the sum of Wilson coefficients and operators, within
the observable ${\cal H}$, does the scale dependence cancels out.
$G_F$ denotes the Fermi coupling constant and the explicit CKM factors
are $\l_q^t = V_{tb}^{} V_{tq}^*$ and $\l_q^u = V_{ub}^{} V_{uq}^*$.
The unitarity relations $\lambda^{c}_q=-\lambda^{t}_q-\lambda^{u}_q$
were already used by us in (\ref{Heff}).

The operators can be chosen as (we only write the most relevant ones):
\bea
\label{oper}
    {\cal O}_2    & = & (\bar{s}_{L}\gamma_{\mu}  c_{L })
                (\bar{c}_{L }\gamma^{\mu} b_{L}) \, , \\
    {\cal O}^u_2    & = & (\bar{s}_{L}\gamma_{\mu}  u_{L })
                (\bar{u}_{L }\gamma^{\mu} b_{L}) \, , \\
    {\cal O}_7    & = & {e}/{g_s^2} m_b (\bar{s}_{L} \sigma^{\mu\nu}
                b_{R}) F_{\mu\nu} \, , \\
    {\cal O}_8    & = & {1}/{g_s} m_b (\bar{s}_{L} \sigma^{\mu\nu}
                T^a b_{R}) G_{\mu\nu}^a \,, \\
    {\cal O}_9    & = & {e^2}/{g_s^2}(\bar{s}_L\gamma_{\mu} b_L)
                \sum_\ell(\bar{\ell}\gamma^{\mu}\ell) \,, \\
    {\cal O}_{10} & = & {e^2}/{g_s^2}(\bar{s}_L\gamma_{\mu} b_L)
                \sum_\ell(\bar{\ell}\gamma^{\mu} \gamma_{5} \ell) \,,
\eea
where the subscripts $L$ and $R$ refer to left- and right- handed
components of the fermion fields. In $b\to s$ transitions the
contributions proportional to $\l_s^u$ are rather small, while in
$b\to d$ decays $\l_d^u$ is of the same order as $\l_d^t$ and they
play an important role in $CP$ and isospin asymmetries.  The operators
${\cal O}_9$ and ${\cal O}_{10}$ only occur in the semileptonic $b
\rightarrow s/d \, \ell^+ \ell^-$ modes.

While the Wilson coefficients $C_i (\m)$ enter both inclusive and
exclusive processes and can be calculated with perturbative methods,
the calculational approaches to the matrix elements of the operators
differ in both cases.  Within inclusive modes, one can use
quark-hadron duality in order to derive a well-defined heavy mass
expansion of the decay rates in powers of $\Lambda_{\hbox{\tiny
    QCD}}/m_b$ (HME).  In particular, it turns out that the decay
width of the $B \rightarrow X_s \gamma$ is well approximated by the
partonic decay rate, which can be calculated in
renormalization-group-improved perturbation theory:
\begin{equation}
\Gamma ( B \rightarrow X_s \gamma) = \Gamma ( b \rightarrow X_s^{parton} \gamma ) +
\Delta^{nonpert.}  
\end{equation}
Non-perturbative effects, $\Delta^{nonpert.}$, are suppressed by
inverse powers of $m_b$ and are well under control thanks to the Heavy
Mass Expansion (HME); they can be further estimated through the
application of the Heavy Quark Effective Theory (HQET).  In exclusive
processes, however, one cannot rely on quark-hadron duality and has to
face the difficult task of estimating matrix elements between meson
states.  A promising approach is the method of QCD-improved
factorization that has recently been systemized for non-leptonic
decays in the heavy quark limit. This method allows for a perturbative
calculation of QCD corrections to naive factorization and is the basis
for the up-to-date predictions for exclusive rare $B$ decays. However,
within this approach, a general, quantitative method to estimate the
important $1/m_b$ corrections to the heavy quark limit is missing.
More recently, a more general quantum field theoretical framework for
the QCD-improved factorization was proposed - known under the name of
Soft Collinear Effective Theory (SCET).

\subsection{Inclusive Modes}
In contrast to the exclusive rare $B$ decays, the inclusive ones are
theoretically clean observables and dominated by the partonic
contributions.  Bound-state effects of the final states are eliminated
by averaging over a specific sum of hadronic states.  Moreover, also
long-distance effects of the initial state are accounted for, through
the heavy mass expansion in which the inclusive decay rate of a heavy
$B$ meson is calculated using an expansion in inverse powers of the
$b$ quark mass.

The optical theorem relates the {inclusive} decay rate of a hadron
$H_b$ to the imaginary part of certain forward scattering amplitudes
\begin{equation}
\Gamma (H_b \rightarrow X) = \frac{1}{2 m_{H_b}} \Im \, \langle 
H_b \mid {\bf T} \mid H_b \rangle\, ,  
\end{equation}
where the transition operator ${\bf T}$ is given by ${\bf T} = i \int
d^4 x \, T [ H_{eff} (x) H_{eff} (0)]$.  It is then possible to
construct an OPE of the operator ${\bf T}$, which gets expressed as a
series of {\it local} operators -- suppressed by powers of the $b$
quark mass and written in terms of the $b$ quark field:
\begin{equation}
{\bf T} \, \, \stackrel{OPE}{=}  \frac{1}{m_b} \big( {\cal O}_0 + \frac{1}{m_b}
 {\cal O}_1 + \frac{1}{m_b^2} {\cal O}_2 + ... \big)\, . 
\end{equation}
This construction is based on the parton--hadron duality, using the
facts that the sum is done over all exclusive final states and that
the energy release in the decay is large with respect to the QCD
scale, $\Lambda_{\hbox{\tiny QCD}} \ll m_b$.  With the help of the
HQET, namely the new heavy-quark spin-flavour symmetries arising in
the heavy quark limit $m_b \rightarrow \infty$, the hadronic matrix
elements within the OPE, $\langle H_b \mid {\cal O}_i \mid H_b
\rangle$, can be further simplified.  The crucial observations within
this well-defined procedure are the following: the free quark model
turns out to be the first term in the constructed expansion in powers
of $1/m_b$ and therefore the dominant contribution. This contribution
can be calculated in perturbative QCD.  Second, in the applications to
inclusive rare $B$ decays one finds no correction of order $1/m_b$ to
the free quark model approximation, and the corrections to the
partonic decay rate start with $1/m_b^2$ only.  The latter fact
implies a rather small numerical impact of the non-perturbative
corrections to the decay rate of inclusive modes.

The $1/m_b^2$ corrections correspond to the OPE for ${\rm T} ({\cal
  O}^\dagger_7 {\cal O}_7)$. There are additional non-perturbative
effects if one also takes into account the operator ${\cal O}_2$. They
can be analysed in a model-independent way and scale with $1/m_c^2$.
Due to small coefficients in the expansion also their impact is very
small \cite{mca,mcb,mcc}.

\subsection{Exclusive Modes}
The naive approach to the computation of exclusive amplitudes consists
in writing the amplitude $A \simeq C_i (\mu_b) \langle {\cal O}_i
(\mu_b) \rangle$ and parametrizing $\langle {\cal O}_i (\mu_b)
\rangle$ in terms of form factors. A substantial improvement is
obtained through the QCD-improved factorization~\cite{BBNS1,BBNS2} and
SCET~\cite{BFL,BFPS,BS,BPS,BCDF,HiNe} approaches.

Let us consider processes involving the decay of a heavy meson into
fast moving light particles ($B\to \gamma e\nu$, $B\to
(\rho,K^*)\gamma$, $B\to K\pi$, ...) and indicate with $Q\sim O(m_b)$
their typical large energy. The idea is to isolate {\it all} the
relevant degrees of freedom necessary to correctly describe the
infrared structure of QCD below the scale $Q$ and associate
independent fields to each of them. It is possible to identify two
distinct {\em perturbative} modes, called hard ($p^2 \sim Q^2$) and
semi-hard ($p^2 \sim \Lambda_{\hbox{\tiny QCD}} Q$). These modes are
produced, for instance, in interactions of energetic light particles
with the heavy quark and the $B$-meson spectator, respectively. These
two modes do not appear in the initial and final states and,
therefore, have to be integrated out. We do not wish to entertain here
a comprehensive discussion of the technicalities involved in this
step. It will suffice to say that the resulting theory (also called
$\hbox{SCET}_{II}$ in the literature) contains only {\em
  non-perturbative} degrees of freedom with virtualities
$O(\Lambda_{\hbox{\tiny QCD}}^2)$ and that hard and semi-hard modes
are reflected in the coefficient functions in front of the operators
of that ($\hbox{SCET}_{II}$) theory.  We note that these coefficients
depend, in general, on energies of order $Q$ and $\Lambda_{\hbox{\tiny
    QCD}}$. Moreover, the hierarchy $\Lambda_{\hbox{\tiny QCD}} \ll Q$
allows for an expansion in the small parameter $\lambda=
\Lambda_{\hbox{\tiny QCD}} /Q$.

Given a process, one has to construct the most general set of
 ($\hbox{SCET}_{II}$)  operators at a given order in $\lambda$, and show
that all the possible gluon exchanges can be reabsorbed, at all orders
in perturbation theory, into form factors and meson light-cone wave
functions. The resulting amplitude is a convolution of these
non-perturbative universal objects with the coefficient functions
encoding the contribution of hard and semi-hard modes. Questions
regarding the convergence of these convolution integrals can be
addressed using symmetries, power counting and dimensional analysis (a
discussion on this point is presented in Ref.~\cite{HiNe}).

The few form factors that describe the transition $B\to M$ (where $M$
denotes a pseudo-scalar or vector meson) can be written
as~\cite{BeFe}:
\bea
F^{B\to M}_i = C_{i} \, \xi^{B\to M} + \phi_B \otimes T_i \otimes \phi_M 
+ O\left({\Lambda \over m_b}\right)\,
\label{ff}
\eea
where $\xi^{B\to M}$ is the so-called non-factorizable (or soft)
contribution to the form factors (actually there is one soft form
factor for the decay into pseudoscalar meson and two for the decay
into vector mesons); $\phi_{B,M}$ are the $B$ and $M$ meson light-cone
wave functions; $C_{i}$ are Wilson coefficients that depend on hard
scales; and $T_i$ are perturbative hard scattering kernels generated
by integrating out hard and semi-hard modes. In Ref.~\cite{BPS2} the
factorization formula Eq.~(\ref{ff}) has been proved at all orders in
perturbation theory and at leading order in $\Lambda_{\hbox{\tiny
    QCD}}/m_b$, using SCET techniques.\footnote{We note that a
  discussion of the convergence of the convolution integrals is still
  missing.} The strength of Eq.~(\ref{ff}) is that it allows us to
express several independent QCD form factors in terms of only one soft
form factor (two in the case of vector mesons) and moments of the
light-cone wave functions of the light pseudo-scalar (vector) and $B$
mesons.

Let us now briefly discuss the form of factorization for the decays
$B\to V \gamma$ (with $V=K^*,\;\rho$). At leading order, only the
operator ${\cal O}_7$ contributes and its matrix element between meson states
is given by an expression similar to (\ref{ff}). The
choice of using either the full QCD form factor $T^{B\to V}$ or the soft
one $\xi_\perp$ clearly is a matter of taste (note that non-perturbative
methods, such as lattice-QCD and light-cone QCD sum rules, give only
informations on the full QCD form factors and not on the soft
contributions alone). The advantage of the QCD-improved factorization
approach is evident in the computation of the next-to-leading order
(in $\alpha_s$) corrections. In fact, one can show that the matrix
elements of the operators ${\cal O}_2$ and ${\cal O}_8$, which  are expected to
contribute at this order, are given by the matrix element of ${\cal O}_7$
times a computable hard scattering kernel. Moreover, spectator
interactions can be computed and are given by convolutions involving
the light-cone wave functions of the $B$ and $V$ mesons.
It must be  mentioned that light-cone wave functions of pseudo-scalar and
vector mesons have been deeply studied using light-cone QCD sum rules
methods~\cite{fily1,fily2,babr,babr2}. On the other hand, not much is
known about the $B$ meson light-cone distribution amplitude, whose
first negative moment enters the factorized amplitude at NLO. Since
this moment enters the factorized expression for the $B\to \gamma$
form factor as well, it might be possible to extract its value from
measurements of decays like $B\to \gamma e \nu$, if it can be shown
that power corrections are under control \cite{daniel}.

Finally, let us stress that a breakdown of factorization is expected
at order $\Lambda_{\hbox{\tiny QCD}}/m_b$~\cite{BBNS2,KaNe,FeMa}. In
Ref.~\cite{KaNe}, in particular, the authors have shown that in the
analysis of $B\to K^* \gamma$ decays at subleading order 
 an infrared divergence is encountered in 
the matrix element of ${\cal O}_8$.
Nevertheless, some very specific power corrections 
might still be computable. Indeed, this is the case for
the annihilation and weak exchange amplitudes in $B\to \rho \gamma$ at
the one-loop level.

\section{$b\to d \gamma$ Transitions}

\subsection{$B\to X_d \gamma$}
Most of the theoretical improvements on the perturbative contributions 
and the power corrections in $1/m_b^2$ and
$1/m_c^2$,
 carried out in the context of the
decay $B \rightarrow X_s \gamma$, can straightforwardly be adapted to 
the decay $B \rightarrow X_d \gamma$; thus, the NLL-improved decay
rate for $B \rightarrow X_d \gamma$ decay has much reduced theoretical
uncertainty \cite{AG7}.  
But as $\lambda^u_d =
V^{}_{ub} V^*_{ud}$ for $b \to d \gamma$ is not small with respect to
$\lambda^t_d = V^{}_{tb} V^*_{td}$ and $\lambda^c_d = V^{}_{cb}
V^*_{cd}$, one also has to take into account the operators
proportional to $\lambda^u_d$ and, moreover, 
the long-distance contributions from the intermediate
$u$-quark in the penguin loops might be important.
However,
there are three {\it soft} arguments that  indicate a small impact of
these non-perturbative contributions: first, one can derive a
model-independent suppression factor $\Lambda_{\hbox{\tiny QCD}} /
m_b$ within these long-distance contributions \cite{Rey}. Second,
model calculations, based on vector meson dominance, also suggest this
conclusion \cite{LDUP}.  Furthermore, estimates of the long-distance
contributions in exclusive decays $B \rightarrow \rho \gamma$ and $ B
\rightarrow \omega \gamma$ in the light-cone sum rule approach do not
exceed 15\% \cite{stollsum}.  Finally, it must be stressed
that there is no spurious enhancement of the form $\log (m_u/\mu_b)$
in the perturbative contribution, as was shown in \cite{GHW,STERMAN}.
All these observations exclude very large long-distance intermediate
$u$-quark contributions in the decay $B \rightarrow X_d \gamma$.
Nevertheless, the theoretical status of the decay $B \to X_d \gamma$
is not as clean as that of $B \rightarrow X_s \gamma$.\\

While the $b \to s$ transitions like $B \to X_s \gamma$ have no
relevant impact on the CKM phenomenology because of  the flatness of the
corresponding unitarity triangle (for example: $V_{ts}$ cannot be
further constrained by the $B \rightarrow X_s \gamma$ measurement if
the unitarity constraint is not used in the theoretical prediction,
because the charm contribution is twice as large as the top
contribution), $b \to d$ transitions give
important complementary information on the  unitarity triangle, which is
also tested by the measurements of $V_{ub}/V_{cb}$, $\Delta
M_{B_d}$, and $\Delta M_{B_d}/ \Delta M_{B_s}$.  Thus, a future
measurement of the $B \rightarrow X_d \gamma$ decay rate will help to
reduce the currently allowed region of the CKM Wolfenstein parameters
$\rho$ and $\eta$ significantly.

Regarding new physics, the branching ratio of $B \rightarrow X_d
\gamma$ might be of interest, because its CKM suppression by the
factor $|V_{td}|^2/|V_{ts}|^2$ in the SM may not be true in extended
models. We also emphasize that in the ratio 
\begin{equation}
\label{dsgamma}
R(d\gamma/s\gamma) \equiv \frac{{\cal B}(B \to X_d \gamma)}
                           {{\cal B}(B \to X_s \gamma)},
\end{equation}
a good part of the theoretical uncertainties cancel out.  It is therefore  
 of particular interest for CKM phenomenology and for the new physics
search.

A measurement of the $B \rightarrow X_d \gamma$ is rather difficult
but perhaps within  the reach of the high-luminosity $B$ factories.  Such a
measurement will rely on high statistics and on powerful methods for
the kaon--pion discrimination.  At present only upper bounds on
corresponding exclusive modes are available (see next section).

The direct {\it normalized} CP asymmetry of the inclusive decay modes
represent another interesting observable:
\begin{equation}
\label{CPdirectdefinition} 
\nonumber 
\alpha_{CP} = \frac{\Gamma(\bar B \rightarrow X_{s/d}\gamma)
     -\Gamma(B \rightarrow  X_{\bar s/\bar d}\gamma)}
     {\Gamma(\bar B \rightarrow  X_{s/d} \gamma)
     +\Gamma(B \rightarrow  X_{\bar s/\bar d}\gamma)}.
\end{equation}
CLEO has already presented a measurement of the CP asymmetry in the
inclusive decay $B \to X_s \gamma$, actually a measurement of a
weighted sum, $\alpha_{CP} = 0.965 \alpha_{CP}(B \rightarrow X_s
\gamma) + 0.02 \alpha_{CP}(B \rightarrow X_d \gamma)$ \cite{CleoCP},
which already excludes very large effects.  The same conclusion can be
deduced from the measurements of the CP asymmetry in the exclusive
modes (see next section). 

Theoretical NLL QCD predictions of the {\it normalized} CP asymmetries
of the inclusive channels (\cite{AG7,KaganNeubert}) within the SM can
be expressed by the approximate formulae (see \cite{SoniWu}):
\begin{equation}
\begin{array}{lll}                   
  \alpha_{CP}({B \rightarrow  X_s \gamma}) &\approx&  0.334 \times
\Im [\epsilon_s] \approx + 0.6 \% \, , \\
  \alpha_{CP}({B \rightarrow  X_d \gamma}) &\approx&  0.334 \times 
\Im [\epsilon_d] \approx - 16 \%,  \label{SMnumbers}
\end{array}
\end{equation}
where
\bea
\epsilon_s = \frac{V_{us}^* V_{ub}}{V_{ts}^*V_{tb}} \simeq 
-\lambda^2( \rho -i \eta), \,     
\epsilon_d = \frac{V_{ud}^*V_{ub}}{V_{td}^*V_{tb}} \simeq
        \frac{ \rho -i \eta}{1- \rho +i \eta} \nonumber
\eea
Numerically, the best-fit values of the CKM parameters are used. 
The two CP asymmetries are connected by the relative factor $\lambda^2
\, ((1-\rho)^2 + \eta^2)$. Moreover, the small SM prediction for the
CP asymmetry in the decay $B \rightarrow X_s \gamma$ is a result of
three suppression factors.  There is an $\alpha_s$ factor needed in
order to have a strong phase; moreover, there is a CKM suppression of
order $\lambda^2$ and there is a GIM suppression of order
$(m_c/m_b)^2$,  reflecting the fact that in the limit $m_c = m_u$ any CP
asymmetry in the SM would vanish.

It will be rather difficult to make an inclusive measurement of the CP
asymmetry in the $b \to d$ channel. However, based on CKM
unitarity, one can derive the following U-spin relation between the
{\it un-normalized} CP asymmetries \cite{Soares}:
\begin{equation} 
\label{resincg1}
\Delta \Gamma (B \to X_s \gamma) +
\Delta \Gamma (B \to X_d \gamma) = 0. 
\end{equation}
Within the inclusive channels, one can rely on parton--hadron duality
and can actually compute the U-spin breaking by keeping a
non-vanishing strange quark mass \cite{mannelhurth}.  Going beyond the
leading partonic contribution, one can further check if the large
suppression factor from the U-spin breaking is still effective, in
addition to the natural suppression factors already present in the
corresponding branching ratios \cite{mannelhurth2}; this finally leads
to the SM prediction
\begin{equation} 
\label{resincg3}
\Delta \Gamma (B \to X_s \gamma) +
\Delta \Gamma (B \to X_d \gamma) = 1 \times 10^{-9}. 
\end{equation}
This prediction provides a very clean SM test,
whether generic new CP phases are active or not.  Any significant
deviation from the estimate would be a direct hint of  non-CKM
contributions to CP violation.

\subsection{$B\to \rho \gamma$}
In the analysis of exclusive $B\to V \gamma$ decays (with $V=K^*$,
$\rho$, $\omega$) 
we will construct the various observables in terms of the 
$CP$-averaged quantities -
which are much easier to measure than the individual channels -
unless otherwise stated. 
In the NLL approximation, this procedure is
equivalent to defining two distinct observables for the charge-conjugate
modes and {\em then} perform the average. 
The ratios $R (\rho \gamma/K^* \gamma)$ are given by
\bea
R^\pm (\rho \gamma/K^ \gamma) =  \left| V_{td} \over V_{ts} \right|^2
 {(M_B^2 - M_\r^2)^3 \over (M_B^2 - M_{K^*}^2)^3 } \z^2
(1 + \D R^\pm) \; , \nn\\
R^0 (\rho \gamma/K^* \gamma) = {1\over 2} \left| V_{td} \over V_{ts} \right|^2
 {(M_B^2 - M_\r^2)^3 \over (M_B^2 - M_{K^*}^2)^3 } \z^2
(1 + \D R^0) \; , \nn
\label{rapp}
\eea
where $\z=\xi_{\perp}^{\rho}(0)/\xi_{\perp}^{K^*}(0)$, and 
$\xi_{\perp}^{\rho}(0)$ and $\xi_{\perp}^{K^*}(0))$ are the form factors
at $q^2=0$ in the effective heavy quark theory for the decays $B \to
\rho (K^*)\gamma$~\cite{bdgAP}. There are several estimates 
of the quantity $\z$ in
the present literature coming from light-cone QCD sum rules
(LCSR)~\cite{Ali:vd}, hybrid LCSR~\cite{narison}, improved
LCSR~\cite{ball} and quark models~\cite{melikhov}.  In the numerical
analysis we adopt the value $\zeta=0.76 \pm 0.10$; the central value is
taken from the LCSR approach while the error is increased in order to
accommodate all the other determinations. 
The quantities $(1 +\Delta R^{\pm,0})$ entail the explicit
$O(\alpha_s)$ corrections as well as the power-suppressed annihilation
contributions proportional to $\lambda^u_d$. The latter contribution,
in particular, is effective only for charged $B$ decays (weak
annihilation); in fact, W-exchange amplitudes are smaller because of
the ratio $Q_d/Q_u = -1/2$ and of colour suppression. The ratio
$R(\rho\gamma/ K^* \gamma)$ acquires, therefore, a tiny dependence on
the CKM angle $\alpha$. Within the SM, the numerical value of these
NLO corrections are $\Delta R^\pm = 0.055 \pm 0.13$ and $\Delta R^0 =
0.015 \pm 0.11$.  Explicit expressions for these quantities, which are
valid in the presence of beyond-the-SM physics, can be found
 in Refs.~\cite{bdgAP,AlLu}.

Further important observables are the isospin breaking ratio given by
\bea
\D (\r\g) &=& 
\frac{\Gamma (B^+ \to \r^+ \g) -
\Gamma (B^- \to \r^- \g) }{2\, (\Gamma (B^0 \to \r^0 \g) + \Gamma (\bar{B}^0
\to \bar{\r}^0 \g)) } -1 
\label{dnlo}
\eea
and the CP asymmetries in the charged and neutral modes, 
\bea
A^\pm_{CP} (\r\g) &=& \frac{\Gamma (B^- \to \r^- \g) -
\Gamma (B^+ \to \r^+ \g) }{\Gamma (B^- \to \r^- \g) + \Gamma (B^+
\to \r^+ \g) } \label{acp} \,, \\
A^0_{CP} (\r\g) &=& 
\frac{
\Gamma (\bar B^0 \to \r^0 \g) - \Gamma (B^0 \to \r^0 \g) 
}{
\Gamma (\bar B^0 \to \r^0 \g) + \Gamma (B^0 \to \r^0 \g) 
} \, .
\eea

Recently, the BABAR collaboration has reported a significant
improvement on the upper limits of the branching ratios for the
decays $B^0(\bar B^0) \to \rho^0\gamma$ and $B^\pm \to \rho^\pm
\gamma$. Averaged over the charge-conjugated modes, the current
$\cl{90}$ upper limits are~\cite{babar:convery}:
\bea
{\cal B}(B^0 \to \rho^0 \gamma) &<& 1.4 \times 10^{-6} \; , \\
{\cal B}(B^\pm \to \rho^\pm \gamma) &<& 2.3 \times 10^{-6} \; , \\
{\cal B}(B^0 \to \omega \gamma) &<& 1.2 \times 10^{-6} \; .
\eea
They have been combined, using isospin weights for $B \to \rho \gamma$
decays and assuming ${\cal B}(B^0 \to \omega \gamma)={\cal B}(B^0 \to
\rho^0 \gamma)$, to yield the improved upper limit
\beq
{\cal B}(B \to \rho \gamma) < 1.9 \times 10^{-6}\; .
\eeq
Note that the equality between the $\rho$ and $\omega$ branching
ratios receives $SU(3)$-breaking corrections that can be as large as
$20\%$. 
The current measurements of the branching ratios for $B \to K^*
\gamma$ decays by BABAR~\cite{babar:grauges},
\bea
{\cal B}(B^0\to K^{*0}\gamma)=(4.23\pm 0.40\pm 0.22)\times 10^{-5}\\
{\cal B}(B^+\to K^{*+}\gamma)=(3.83\pm 0.62\pm 0.22)\times 10^{-5}
\eea
are then used to set a $\cl{90}$ upper limit on the ratio of the
branching ratios~\cite{babar:convery}:
\bea
& & R(\rho \gamma/K^*\gamma) \equiv  \frac{{\cal B}(B \to \rho \gamma)}{{\cal B}(B
\to K^* \gamma)} < 0.047 \; .
\eea
This bound is typically a factor 2 away from the SM
estimates~\cite{bdgAP}, which we quantify more precisely in this
letter. In beyond-the-SM scenarios, this bound provides a highly
significant constraint on the relative strengths of the $b \to d
\gamma$ and $b \to s \gamma$ transitions.

\begin{figure}[!t]
\hbox to\hsize{\hss
\includegraphics[width=\hsize]{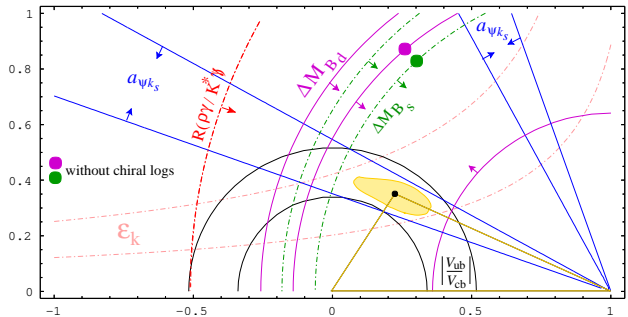}
\hss}
\caption[1]{\it Unitary triangle fit in the SM and the resulting 95\%
C.L. contour in the $\bar \rho$ - $\bar \eta$ plane. 
The solid lines show the upper bounds (with and without chiral logs) 
due to  $\Delta M_{B_d}$, the dot-dashed lines the ones due to $\Delta M_{B_s}$.
The impact of the
$R(\r\g/K^*\g) < 0.047$ constraint is also shown.} 
\label{fig:utsm}
\end{figure}

Let us present an updated analysis of the constraints in the $(\bar
\r, \bar \et)$ plane from the unitarity of the CKM matrix, including
the measurements of the CP asymmetry $a_{\psi K_s}$ in the decays
$B^0/\overline{B^0} \to J/\psi K_s$ (and related modes), and show the
impact of the upper limit $R(\rho \gamma/K^* \gamma) \leq
0.047$~\cite{babar:convery}.  The SM expressions for $\epsilon_K$
(CP-violating parameter in $K$ decays), $\D M_{B_d}$ ($B_d^0$--$\bar
B_d^0$ mass difference), $\D M_{B_s}$ ($B_s^0$--$\bar B_s^0$ mass
difference) and $a_{\psi K_s}$ are fairly standard and can be found,
for instance, in \rf{utAL}; the values of the theoretical parameters
and experimental measurements that we use are taken from
Ref.~\cite{AlLu}.
The SM fit of the unitarity triangle is presented in \fig{fig:utsm}.
Note that where  the hadronic parameters $f_{B_d} \sqrt{\hat B_{B_d}}$
and $\zeta_s$ are concerned, we adopt very 
recent lattice estimates that take into account
uncertainties induced by the so-called chiral logarithms~\cite{lellouch}.
These errors are extremely asymmetric and, once taken into account, reduce
sizeably the impact of the $\D M_{B_s}/ \D M_{B_d}$  lower bound on the
UT analysis. In Fig.~\ref{fig:utsm} we explicitly show what happens to
the allowed regions once these errors are taken into account. The
$\cl{95}$ contour is drawn, taking into account chiral logarithms uncertainties.

As the bound from the current upper limit on $R(\rho \gamma/K^*
\gamma)$ is not yet competitive to the ones from either the
measurement of $\Delta M_{B_d}$ or the current bound on $\Delta
M_{B_s}$, we use the allowed $\bar \r - \bar \et$ region to
work out the SM predictions for the observables in the radiative
$B$-decays described above.  Taking into account these errors and the
uncertainties on the theoretical parameters, we find the following SM
expectations for the radiative decays:
\bea
R^\pm (\r \g /K^* \g) &=& 0.023 \pm 0.012 \; ,\\
R^0 (\r \g /K^* \g) &=& 0.011\pm 0.006 \; ,\\
\D (\r \g ) &=& 0.04^{+0.14}_{-0.07} \; ,\\
A_{CP}^\pm (\r\g) &=& -0.10^{+0.02}_{-0.03} \; , \\
A_{CP}^0 (\r\g) &=& -0.06 \pm 0.02 \; .
\eea
In the CP asymmetries the uncertainties due to formfactors 
cancel out to a large extent, however, the scale dependence 
is rather large because the CP asymmetries arise at the 
$O(\alpha_s)$. 
It is interesting to work out the extremal values of $R(\rho
\gamma/K^* \gamma)$ compatible with the SM UT analysis. 
Any measurement of $R(\rho \gamma/K^* \gamma)$ whose central 
value lies in the range $(0.013,0.037)$ would
be compatible with the SM, irrespective of the size of the
experimental error. The error induced by the imprecise determination
of the isospin breaking parameter $\z$ currently limits the
possibility of having a very sharp impact from $R(\rho
\gamma/K^*\gamma)$ on the UT analysis.

\begin{figure}[!t]
\hbox to\hsize{\hss
\includegraphics[width=\hsize]{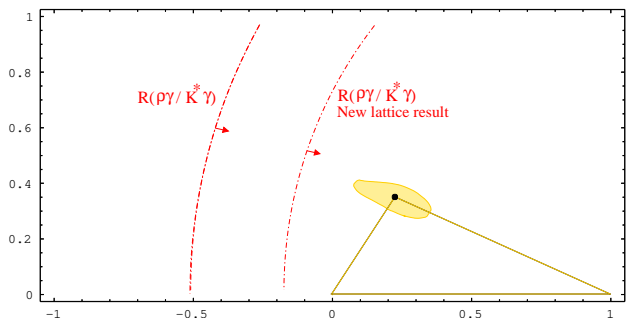}
\hss}
\caption[1]{ \it The impact of the new lattice QCD estimate of the 
ratio $\zeta$ on the $R(\r\g/K^*\g)$ constraint.}
\label{fig:rappsm} 
\end{figure}

Let us comment on the impact of a recent preliminary lattice
determination of the $B\to (K^*,\rho )$ form factor at zero
recoil~\cite{damir}: $F^{K^*} (0) = 0.25(5)(2)$ and $\zeta = 0.91
(8)$. In the first place, note that the central value of the $K^*$
form factor is in perfect agreement with the indirect determinations
obtained in Refs.~\cite{BeFeSe,BoBu,bdgAP}.  In these papers, the
authors compute $BR(B\to K^* \gamma)$ at NLL, using the inclusive
channel $B\to X_s \gamma$ to extract the value of the Wilson
coefficient $C_7$. Their conclusion is that in order to accommodate
the experimental data, the form factor at zero-recoil has to be
substantially smaller than the typical light-cone QCD estimate
($T^{K^*}_{s.r.} (0) = 0.38\pm 0.06$). For instance, in
Ref.~\cite{bdgAP}, the fitted value of the form factor is $T^{K^*}(0)
= 0.27 \pm 0.04$. In second place, the higher central value for the
ratio $\zeta$ given in Ref.~\cite{damir} strengthens the impact of the
current $R(\rho\gamma/K^*\gamma)$ upper limit. In
Fig.~\ref{fig:rappsm}, the additional line has been obtained using
this new determination of the form factors ratio and gives a bound
comparable to the $\Delta m_{B_s}$ one. In order to fully trust this
new lattice estimate of the ratio $\zeta$, an independent cross-check
of this result is mandatory; moreover it is necessary to analyse the
old QCD sum rules estimate deeper so as to understand the reasons of
this discrepancy.

Let us finally discuss  the  analysis of 
the exclusive modes in supersymmetric
models and 
entertain two variants of the MSSM called in the literature
 MFV~\cite{mfvCDGG} and Extended-MFV~\cite{emfvAL} models. In MFV
models, all the flavour changing sources other than the CKM matrix are
neglected. 
In this class of models there are essentially no additional
contributions (on top of the SM ones) to $a_{\psi K_S}$ and $\D
M_{B_s}/\D M_{B_d}$, while the impact on $\epsilon_K$, $\D M_{B_d}$
and $\D M_{B_s}$ is described by a single parameter, $f$, whose value
depends on the parameters of the supersymmetric models~\cite{utAL}.
EMFV models are based on the assumption that all the superpartners are
heavier than 1 TeV with the exception of the lightest stop; no
constraints are imposed on the off diagonal structure of the soft
breaking terms. It can be shown~\cite{emfvAL} that under these
assumptions there are only two new parameters in addition to the MFV
ones, namely: $\d_{\tilde u_L \tilde t} = M^2_{\tilde u_L \tilde t} /
(M_{\tilde t} M_{\tilde q}) \times V_{td}/|V_{td}|$ and $\d_{\tilde
  c_L \tilde t} = M^2_{\tilde c_L \tilde t} / (M_{\tilde t} M_{\tilde
  q}) \times V_{ts}/|V_{ts}|$. Where $\tilde t$ is the lightest stop
mass eigenstate and $M^2$ is the up-squark mass matrix given in a
basis obtained from the SCKM one after the diagonalization of the
$2\times 2$ stop submatrix. Since we are interested in the
phenomenology of $b\to d$ transitions, we will consider here only
$\d_{\tilde u_L \tilde t}$. With the inclusion of this new parameter,
the description of the UT-related observables needs one more complex
parameter, $g = g_R + i g_I$~\cite{emfvAL}. A signature of these
models is the presence of a new phase in the $B_d^0-\bar B_d^0$ mixing
amplitude. Using the parametrization $M_{12}^d = r_d^2 e^{2 i
  \theta_d} M_{12}^{\rm SM}$, we get $r_d^2 = |1+f+g|$ and $\theta_d =
1/2 \arg (1 + f + g)$. This implies new supersymmetric contributions
to the CP asymmetry $a_{\psi K_s}$.

The phenomenology of the MFV and EMFV models, analysed by  
scatter plots over the supersymmetric parameter space, shows the  
discrimation power of exclusive modes, if one focus on ratios of exclusive
observables and their correlation.
\begin{figure}[!t]
\hbox to\hsize{\hss
\includegraphics[width=\hsize]{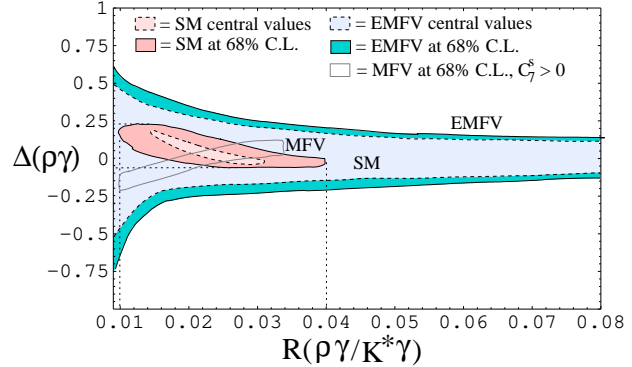}
\hss}
\caption[1]{ \it Correlation between $R(\r\g/K^*\g)$ and
$\Delta(\rho\gamma)$ in the SM and in MFV and EMFV models. The
light-shaded regions are obtained varying $\bar \rho$, $\bar \eta$,
the supersymmetric parameters (for the MFV and EMFV models) and using
the central values of all the hadronic quantities. The darker regions
show the effect of $\pm 1 \sigma$ variation of the hadronic
parameters.}
\label{fig:rapp_delta}
\end{figure}
If one also  scan over $\bar \rho$ and $\bar \eta$, and require that each
point satisfy the bounds that come from direct searches, from the
$B\to X_s \g$ branching ratio, and from the UT-related observables, one 
finally finds the 
surviving regions presented in Fig.~\ref{fig:rapp_delta}.
It shows the correlation of the isospin breaking ratio $\Delta(\rho\gamma)$
and the ratio of the branching ratios
$R(\rho\gamma/K^*\gamma)$.
The
light-shaded regions are obtained using the central values of the
input parameters while the dark-shaded ones result from the inclusion
of their $1 \sigma$  errors. 
In the
MFV case, there are two distinct regions that correspond to the
negative (SM-like) and positive $C_7^s$ case. For $C_7^s<0$, the
allowed regions in MFV almost coincide with the SM ones and we do not
draw them. For $C_7^s>0$, the allowed regions are different and, in
general, a change of sign of both the CP-asymmetries (compared to the
SM) is expected. We note that the latter scenario needs very large
SUSY contributions to $C_7^s$, arising from the chargino-stop
diagrams, and for fixed values of $\tan \b_S$ it is possible to set an
upper limit on the mass of the lightest stop squark.

\section{$b\to s \ell^+ \ell^-$ Transitions}

\subsection{$B\to X_s \ell^+ \ell^- $}

In comparison to  the 
$B \rightarrow X_s \gamma$, the inclusive $ B \rightarrow X_s \ell^+\ell^-$ 
decay presents a complementary and also more complex test of the SM. 
This decay  is  dominated 
by perturbative  contributions if the  $c \bar c$ resonances 
that show up as large peaks in the dilepton invariant mass spectrum 
are removed by appropriate  kinematic cuts. 
In the 'perturbative windows', namely  
in the low-$\hat{s}$ region $ 0.05 <  \hat{s} = q^2 / m_b^2 < 0.25 $ and 
also in the high-$\hat{s}$ region with $ 0.65 < \hat{s}$,
theoretical predictions for the invariant mass spectrum
are dominated by the purely perturbative contributions, 
and a theoretical precision comparable with  the one reached  
in the decay $B \rightarrow X_s \gamma$ is in principle possible. 
Regarding the choice of precise cuts in the dilepton mass 
spectrum, it is important that  one directly compares theory and  
experiment using the same energy cuts and avoids any kind of
extrapolation. 

In the high-$\hat{s}$ region, one should encounter the breakdown of the 
heavy mass expansion at the endpoint.
Integrated quantities are still defined; nevertheless   
one finds sizeable $\Lambda^2_{QCD}/m_b^2$ non-perturbative 
corrections within this region.

The decay $B \rightarrow X_s \ell^+\ell^-$ is particularly 
attractive because of 
kinematic observables such as 
the invariant dilepton mass spectrum and the forward--backward 
(FB) asymmetry. They are usually normalized by the semi-leptonic
decay rate in order to reduce the uncertainties due 
to bottom quark mass and CKM angles and are defined as follows:
\begin{equation}\label{decayamplitude}
R^{\ell^+\ell^-}_{quark}(\hat{s})= \frac{d}{d \hat{s}}\Gamma( b \to X_s\ell^+\ell^-) /
 \Gamma( b \to X_c e\bar{\nu}),
\end{equation}
\begin{eqnarray}\label{forwardbackward}
A_{FB}(\hat{s}) & =& \frac{1}{\Gamma( b \to X_ce\bar{\nu})}  \nonumber\\ 
& & \hskip -1.5cm \times \int_{-1}^1 d\cos\theta_\ell ~
\frac{d^2 \Gamma( b \to X_s \ell^+\ell^-)}{d \hat{s} ~ d\cos\theta_\ell}
\mbox{sgn}(\cos\theta_\ell), 
\end{eqnarray}
here $\theta_\ell$ is the angle between $\ell^+$ and $B$ momenta in
the dilepton centre-of-mass frame.  The so-called `normalized' FB
asymmetry, which is also often used, is given by
\begin{eqnarray}
\overline{A}_{FB}(\hat{s}) =
{\displaystyle \int_{-1}^1 d\cos\theta_\ell ~
 \frac{d^2 \Gamma( B\to X_s \ell^+\ell^-)}{d \hat{s} ~ d\cos\theta_\ell}
\mbox{sgn}(\cos\theta_\ell) \over \displaystyle 
 \int_{-1}^1 d\cos\theta_\ell ~
 \frac{d^2 \Gamma( B\to X_s \ell^+\ell^-)}{d \hat{s} ~ d\cos\theta_\ell}} \,.
\hskip -2cm \nn\\ 
\end{eqnarray}

For the low-$\hat s$ region the present partonic NNLL prediction is
given by (see \cite{Asa1,Adrian2,MISIAKBOBETH}):
\begin{equation}
\int_{0.05}^{0.25}  \, d \hat{s} \,  R^{\ell^+ \ell^-}_{quark}(\hat{s}) 
\, = \, (1.27 \pm 0.08_{scale} \,)\, \times  10^{-5} 
\label{partonicfinal}
\end{equation}
The error quoted in (\ref{partonicfinal}) 
reflects only the renormalization  scale uncertainty and is
purely perturbative.  
There is no additional problem due to the charm mass renormalization
scheme ambiguity within the decay $B \rightarrow X_s \ell^+ \ell^-$ 
because the charm dependence starts already at one loop, in contrast 
to the case of the decay $B \rightarrow X_s \gamma$. 
The charm dependence itself leads to an additional uncertainty
of $\sim 7.6 \%$ within the partonic quantity (\ref{partonicfinal}), 
if the pole mass is varied,  $m_c^{pole}/m_b^{pole} = 0.29 \pm 0.02$.   

The impact of the NNLL contributions
is significant. The large matching scale $\mu_W$ uncertainty of $16 \%$ 
of the NLL result was removed; the low-scale uncertainty $\mu_b$ 
of $13 \%$ was cut in half; and also the central value of the integrated
low dilepton spectrum (\ref{partonicfinal})   
was significantly changed by more than $10 \%$ due to NNLL corrections. 
Using the measured semi-leptonic branching ratio ${\cal B}^{sl}_{exp.}$,
the prediction for the corresponding  branching ratio 
is given by 
\begin{eqnarray}
&& {\cal B}(B \rightarrow X_s \ell^+ \ell^-)_{\hbox{Cut:}\,\,\hat{s}  
\in [0.05,0.25]} = \\
&=&\, {\cal  B}^{sl}_{exp.} \, \int_{0.05}^{0.25}  \, d \hat{s} \, \big[ R^{\ell^+ \ell^-}_{quark}(\hat{s}) + R_{m_b^2}(\hat{s}) + R_{m_c^2}(\hat{s})  \big]
\nonumber\\
&=&\, (1.36 \pm 0.08_{scale} \,)  \times 10^{-6} \nonumber
\label{finalll}
\end{eqnarray}
$R_{m_b^2}(\hat{s})$ and $R_{m_c^2}(\hat{s})$ 
are the non-perturbative contributions scaling with 
${1/m_b^2}$ and ${1/m_c^2}$.
The recent first measurement of BELLE, with a rather large
uncertainty \cite{BELLEbsll2}, is compatible with this SM prediction.

\begin{figure}[!t]
\hbox to\hsize{\hss
\includegraphics[width=0.85\hsize]{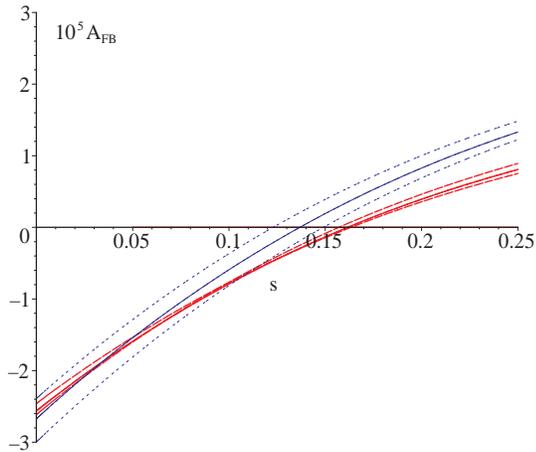}
\hss}
\caption[1]{ \it Comparison between NNLL and  NLL results for 
$A_{FB}(s)$ in the low $s$ region. 
The three thick lines are the NNLL  
predictions for $\mu=5$~GeV (full),
and $\mu=2.5$ and 10 GeV (dashed); the dotted  curves 
are the corresponding NLL results. All curves for $m_c/m_b=0.29$.}
\label{fig:AFB} 
\end{figure}

The phenomenological impact of the NNLL 
contributions on the FB asymmetry is 
also significant \cite{Adrian1,Asa2}. 
The position of the zero of the FB asymmetry, defined by 
$A_{FB}(\hat{s}_0)=0$, is particularly interesting to determine 
relative sign and  magnitude of the Wilson coefficients 
$C_7$ and $C_9$ and it is therefore extremely sensitive to 
possible new physics effects.
The previous NLL result,
where the error is determined by the  
scale dependence, is now modified by 
the  NNLL contributions \cite{Adrian1,Asa2}:
\begin{equation}
 \hat{s}^{NLL}_0 = 0.14 \pm 0.02~,\,\,\,\,\hat{s}^{NNLL}_0 = 0.162 \pm 0.008~.
\label{eq:s0NLO}
\end{equation}
  In the NNLL  case the variation of the result induced by 
the scale dependence is 
accidentally very small 
(about $\pm 1\%$) and cannot be 
regarded as a good estimate of missing higher-order effects. 
Taking into account the separate scale variation of both  Wilson 
coefficients $C_9$ and  $C_7$,
and the charm-mass dependence, one estimates a conservative 
overall error on $\hat{s}_0$ of about $5 \%$ \cite{Adrian1}. 
In this $\hat s$ region the non-perturbative  
$1/m_b^2$ and $1/m_c^2$ corrections to $A_{FB}$ 
are very small and also under control. 
An illustration of the shift of the 
central value and the reduced scale dependence 
between NNL and NNLL expressions of $A_{FB}(s)$, in the low-$\hat s$
region, is presented in fig.~\ref{fig:AFB}. 
The complete effect of NNLL
contributions to the FB asymmetry  adds up to a $16\%$ shift compared 
with the  NLL result, with a residual error due to higher-order terms  
reduced at the 5\% level. 
Thus, the zero of the FB asymmetry in the inclusive mode turns out 
to be one of the most sensitive tests for new physics beyond the SM.  

\begin{figure}[!t]
\hbox to\hsize{\hss
\includegraphics[width=\hsize]{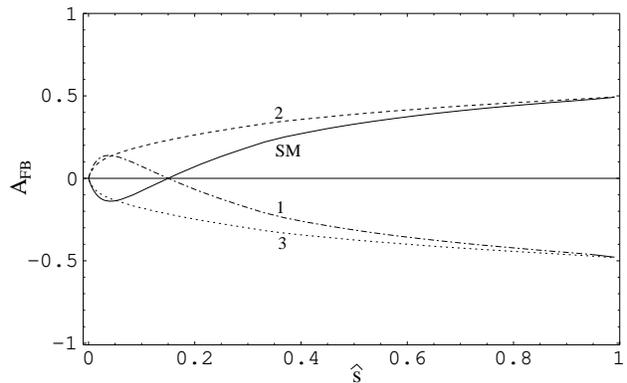}
\hss}
\caption[1]{ \it Four different shapes of the normalized FB  asymmetry 
$\overline{A}_{FB}$
for the decay $B\to X_s \ell^+ \ell^-$. The four curves correspond to four
sample points of the Wilson coefficients that are compatible with
the present measurements of the integrated branching ratios.}
\label{Fig33} 
\end{figure}

The $B$ factories will soon provide statistics and resolution
needed for the measurements of $B \rightarrow X_s \ell^+\ell^-$
kinematic distributions. 
Correspondingly, 
the recently calculated new (NNLL) contributions 
\cite{Asa1,Adrian2,Adrian1,Asa2,MISIAKBOBETH}\footnote{We add here
that  the three-loop mixing is fully are  under control.
A quite recent calculation of a missing NNLL mixing piece leads to 
a correction below $2\%$ \cite{Paolonew}}
have significantly 
improved the sensitivity of the inclusive $B \rightarrow X_s \ell^+ \ell^-$ 
decay in  testing extensions of the SM in the sector of flavour 
dynamics. 
However, with the 
present experimental knowledge the decay $B \rightarrow 
X_s \gamma$ still leads to  the most restrictive constraints
as was found in \cite{AGHL}. 
Especially, the MFV scenarios are already highly constrained and only
small deviations to the SM rates and distributions are possible;
therefore no useful additional bounds from the semi-leptonic modes
beyond what are already known from the $B  \rightarrow  X_s \gamma$
can be deduced for the MFV models at the moment. 
Within the model-independent analysis, the impact of the partial 
NNLL contributions on the allowed ranges for the Wilson coefficients 
was already found to be significant. 
In this analysis, however, only the integrated branching ratios were used 
to derive constraints. It is clear that one needs measurements of the
kinematic distributions of the $B \rightarrow X_s \ell^+ \ell^-$,
the dilepton mass spectrum and the FB asymmetry in order to 
                                   determine the exact values and signs 
of the Wilson coefficients. 
In fig. \ref{Fig33}, the impact of these future measurements 
is illustrated. It shows the shape of the FB asymmetry for the SM and three
additional sample points, which are all still allowed by the present
measurements of the branching ratios; thus, 
even rather rough measurements of the FB asymmetry will either rule out 
large parts of the parameter space of extended  models or show
clear evidence for new physics beyond the SM. 
\subsection{$B\to K^{(*)} \ell^+ \ell^-  $}
For completeness, let us briefly comment on the impact of exclusive
$B\to K^{(*)} \ell^+\ell^-$ modes. First of all, let us stress that
hadronic uncertainties on these exclusive rates are dominated by the
errors on form factors and are much larger than in the corresponding
inclusive decays. In fact, following the analysis presented in
Ref.~\cite{AGHL}, we see that inclusive modes already put much
stronger constraints on the various Wilson coefficients.

Concerning the measurement of a zero in
the spectrum of the forward-backward asymmetry, things are different.  
According to
Refs.~\cite{ABHH,BeFeSe} the value of the dilepton invariant mass
($q^2_0$), for which the differential forward--backward asymmetry
vanishes, can be predicted in quite a clean way. In the QCD
factorization approach at leading order in $\Lambda_{\hbox{\tiny
    QCD}}/m_b$, the value of $q_0^2$ is free from hadronic
uncertainties at order $\alpha_s^0$ (a dependence on the soft form
factor $\xi_\perp$ and the light cone wave functions of the $B$ and
$K^*$ mesons appear at NLL). Within the SM, the authors of
Ref.~\cite{BeFeSe} find: $q_0^2 = (4.1 \pm 0.6) \, \hbox{GeV}^2$. 
As in the inclusive case, 
such a 
measurement will have a huge phenomenological impact.

\section*{Acknowledgements} 
We thank Thorsten Feldmann for a careful reading of the manuscript 
and for useful discussions. 
The work is partially supported by the Swiss National Foundation
and by the EC-Contract HPRN-CT-2002-00311 (EURIDICE).

\end{document}